\documentclass{tlp}
  \title[Theory and Practice of Logic Programming]
        {Multi-threading and Message Communication in Qu-Prolog }

  \author[K. Clark, P.J. Robinson, R. Hagen]
	{KEITH CLARK\\
         Dept. of Computing\\
         Imperial College, London\\
         \email{klc@doc.ic.ac.uk}
         \and 
         PETER J. ROBINSON, RICHARD HAGEN\\
	 Software Verification Research Centre,The University of Queensland, Austraila\\
         \email{pjr@csee.uq.edu.au}}

\jdate{March 2000}
\pubyear{2000}
\pagerange{\pageref{firstpage}--\pageref{lastpage}}

\begin{document}

\label{firstpage}

\maketitle

\begin{abstract}
This paper presents the multi-threading and internet message communication
capabilities of Qu-Prolog. Message addresses are symbolic and the
communications package provides high-level support that
completely hides details of IP addresses and port numbers as
well as the underlying  TCP/IP transport layer.
The combination of the multi-threads and the high level inter-thread 
message communications provide
simple, powerful support for implementing internet distributed intelligent
applications. 
\end{abstract}

\section{Introduction}

Qu-Prolog \cite{QPUM,QPRM} is an extension of Prolog designed primarily as
an implementation and tactic language for interactive theorem provers,
particularly those that carry out schematic proofs.
Qu-Prolog has built-in support
for the kinds of data structures typically encountered in theorem
proving activities such as object variables, substitutions and
quantified terms.
Qu-Prolog is the implementation language of the Ergo theorem prover \cite{ERGO},
which has seen substantial use in development of verified software,
both directly \cite{VSOFT} and indirectly through prototyping
a program refinement tool \cite{PREF}.

As part of our ongoing efforts to scale up our formal development tools,
we are interested in developing multi-user versions of these tools.
In particular, we are interested in implementing a multi-threaded, multi-user
version of Ergo where a collection of people and automated theorem provers 
can work together to produce a proof. 

As a preliminary step to this, we have augmented Qu-Prolog to support
multi-threading and high-level inter-thread communication 
between Qu-Prolog threads running anywhere on the internet.
An initial case study on multi-threaded theorem proving is described in
\cite{MTTP}.

This paper reports on these new features of Qu-Prolog. 
Our main aim in designing the thread mechanism  
is to provide simple and powerful methods for
programming distributed Prolog-based agent applications. 
Some simple examples of using our approach for DAI applications
are given in \cite{DAI}.

Apart from the ability to give symbolic names to threads and the
management of messages, the implementation
of threads is similar to that of other multi-threaded Prologs,
such as BinProlog \cite{BP} and SICStus MT \cite{SICSMT}.
The novelty of the Qu-Prolog approach to threads is the
high level inter-thread message-based communication which
transparently communicates messages independently of
location. 
Inter-thread communication uses the API of McCabe's
{\it InterAgent Communications Model} (ICM) \cite{Mc}.
This means that Qu-Prolog applications can transparently link 
with other applications that use the ICM such as April \cite{ClMc}
applications.

Qu-Prolog inter-thread communication borrows ideas
from both Erlang \cite{Erlang} and April.
Threads in Qu-Prolog behave as communicating processes which have a
single message buffer of unread messages. 
They will suspend if they want to read a message and the buffer is empty, 
and will then resume as soon as a message, 
communicated from another thread or application, is added to the buffer.

The organization of the paper is as follows. In Section 2 we 
briefly describe Qu-Prolog threads. In Section 3 we present the 
high-level inter-thread communication of Qu-Prolog and in Section 4 we discuss its 
implementation.
Sections 5 and 6 illustrate the use of threads and high-level communication
by presenting an implementation of the Linda model for interprocess 
communication and an implementation of distributed  query processing 
in which each Prolog query server can process many queries 
simultaneously.
In Section 7 we compare Qu-Prolog with SICStus-MT, BinProlog, CIAO,
Mozart-Oz, Erlang and April.

\section{Threads}

Qu-Prolog is implemented as an extended WAM emulated in C++. 
Thread execution is controlled by a scheduler that is responsible for
time-slicing threads, managing blocking of I/O and ICM message and signal
handling.

In the implementation the threads within a single Qu-Prolog process 
share the static code
area and the asserted and recorded databases.  
On the other hand, threads carry out independent Qu-Prolog computations
and so, for example, have separate heaps, stacks and trails. 

The Qu-Prolog thread library contains predicates for creating and deleting
threads, symbolically naming threads, and for controlling thread execution.
The sizes of the WAM data areas for each thread can be set at creation time so
that the size of each individual thread can be tailored to its intended
use.

The predicates
\begin{verbatim}
thread_fork_anonymous(-ThreadID, +Goal, +Sizes)
thread_fork(-ThreadID, +Name, +Goal, +Sizes)
\end{verbatim}
create a new thread and execute \verb~Goal~ within the thread.
When the goal finishes executing (by success or failure) the thread
terminates. The sizes of the various data areas, such as the heap, is specified
in the \verb~Sizes~ data structure. If this argument is missing, the default sizes are used.
\verb~ThreadID~ is the ID of the created thread, and in the second predicate,
\verb~Name~ is the symbolic name given to the thread.

Typically, the main or initial thread of an application is in charge of forking
threads. If this is an `intelligent' server application this initial thread 
is usually programmed as a top level tail recursive or repeat/fail loop that
responds to messages sent to it from any number of client applications. 
We shall give
an example of this later. 
For some of these `queries' the main thread may fork one or more new
threads to process the `query' or to engage in a conversation with the
client. The client, in turn, can fork a thread for each conversation.  
This  enables peer
to peer, or agent to agent application programming, rather than just
client/server programming.

The predicates
\begin{verbatim}
thread_forbid
thread_resume
\end{verbatim}
are used to allow a thread to take control so that it can, for example, perform
an atomic operation like an assert on the shared dynamic clauses. \verb~thread_forbid~ prevents
other threads from having a time-slice and \verb~thread_resume~ resumes 
time-slicing.

Apart from being able to symbolically name threads, the implementation
of threads is similar to that of other multi-threaded Prologs and will
not be discussed further in this paper, except where it relates to 
communication.

\section{Inter-Thread Communications}

In this section we give a user-level view of inter-thread communication
and symbolic addressing in Qu-Prolog. By inter-thread communication we mean
communication between any two threads whether they are in the same
Qu-Prolog process or
different Qu-Prolog processes, even on different machines.

Throughout this section we concentrate on the higher-level communication
support in Qu-Prolog. We consider two layers: the basic support for
inter-thread communication; and a very-high-level layer built on the
basic support.
Qu-Prolog also supports low-level communications via sockets but this is
reasonably standard and is therefore not discussed in this paper. 
The socket level primitives can be used for communicating with pre-existing internet services, such as HTTP or FTP servers.

From a user perspective, the higher-level communication support treats
communication between threads in a uniform way - 
all messages to a thread, irrespective of the sources of the messages, are
added to the end of the thread's single message buffer.

Each message that appears in a thread's message buffer has three
components: the actual message; the sender address; and the reply-to address.
Generally, the reply-to address is the same as the sender address (the default)
but the sender can set the reply-to address when, for example,
the message is forwarded.

Each thread address also has three components: the identity of the thread; the 
identity of the Qu-Prolog process that contains the thread; and the
identity of the machine running the Qu-Prolog
process\footnote{Strictly speaking this is the identity of the host
running the ICM communications server with which the Qu-Prolog
process is registered. This is usually the machine on which the
Qu-Prolog process is running, but need not be. See section 4.}.

For threads, the identity is either an integer representing the
thread ID given to the thread at creation time or its symbolic name given to 
it when it was created or by the use of the 
predicate \verb~thread_set_symbol/1~.

The identity of a process is a symbolic name given to the process
when it is started at the operating system level. The machine identity is
typically the IP 
address, but can 
also be any of its internet symbolic names.

\subsection{Basic inter-thread communication}
 The basic communication layer provides support for constructing
address data structures, for sending messages and for accessing
the message buffer.

The most general message send predicate is
\begin{verbatim}
ipc_send(+Message, +ToAddress, +ReplyAddress, +Options)
\end{verbatim}
where \verb~Message~ is a Prolog term that is the message,
\verb~ToAddress~ is an address data structure representing the
message destination, \verb~ReplyAddress~ is an address data structure 
representing the reply-to address, and \verb~Options~ is a list
of flags that control how the message term is represented as a string of
characters.

The option flags are \verb~remember_names~ and \verb~encode~. 
One important requirement of Qu-Prolog is to support interaction with
symbolic data. This is achieved by being able to associate variable names 
with variables. Qu-Prolog has variants of the read and write predicates
that remember variable names that are input and generate names for unnamed 
variables on output. An important consequence of this is that if a variable
is read in whose name is the same as an existing variable then the
input variable will become the existing variable.

This feature is extended to messages. 
If the \verb~remember_names~ option is set then any unnamed variables in 
the message are given names.
This option, in combination with the equivalent option for receiving messages,
provides a variable connection across threads and thereby supports
processing of schematic data across threads.  
In particular, if one thread T1 sends a sequence of messages to the same
destination thread T2, some or all of which contain the same internal variable
{\tt X1} of T1,
this will be given the same name {\tt N} in all the messages in which it
occurs. T2 will map each occurrence of {\tt N},
in the different messages, into the same internal variable {\tt
X2} of T2. This allows incremental transmission of queries between
threads. 

The \verb~encode~ option determines if efficient Prolog term compression
is to be used or if the term is to be sent in `raw' string form.
For Qu-Prolog-to-Qu-Prolog communication, the encoded form is much more
efficient both for writing and reading. If, however, a message is being
sent to a non-Qu-Prolog process then the raw string form is usually
more appropriate.

There are three predicates for accessing the message buffer. They are
(in their most general forms)
\begin{verbatim}
ipc_recv(?Message, ?FromAddress, ?ReplyAddress, +Options)
ipc_peek(?Message, -Reference, ?FromAddress, ?ReplyAddress, +Options)
ipc_commit(+Reference)
\end{verbatim}

\verb~ipc_recv~ reads the first message in the buffer and unifies
the message and its address data structures with the corresponding
supplied arguments.
It fails if the unification with this first message fails. So, usually
all the arguments of the call are unbound variables.  

\verb~ipc_peek~ searches the message buffer (from the beginning) for
a message that unifies with the supplied message and address arguments,
returning \verb~Reference~ as a reference to the matching message
in the buffer (really a pointer to the message buffer). 
On backtracking it will try to find another match.

\verb~ipc_commit~ is used to remove a message from the message buffer.
Typically, \verb~ipc_peek~ and \verb~ipc_commit~ are used in combination to
search for a particular message and then remove it.

For both \verb~ipc_recv~ and \verb~ipc_peek~, if no (matching) message is found then the behaviour
of the call is determined by the \verb~timeout~ flag in the options list.
If the flag is set to \verb~block~ then the call is delayed until
another message arrives (the default). If the flag is \verb~poll~
then  the call fails immediately. Otherwise, if the flag is an integer
$n$ then the call will suspend for up to $n$ seconds. If no message arrives
in that time then the call fails.

The other possible option is \verb~remember_names~.
If this is set then a connection will be made between named variables in the
message and corresponding named variables in the thread, which
typically were variables of previous messages received using this option.
If the option is not set then no connection is made between variables
in the message and variables of the thread.  This 
provides `separation of variables'.

The encode option is not required because this information is part of the
incoming message and is used to determine if decoding is required.

\subsection{High-level inter-thread communication}
The higher level layer provides application writers with a powerful yet 
simple interface to the basic inter-thread communications layer. 
Again, there are
two parts to this layer: management of addresses, and communication.

In this layer full addresses take the form
\begin{verbatim}
ThreadName:ApplicationName@HostName.
\end{verbatim}

As with
email communication, the global name can be shortened for local
communications. Just \verb~ThreadName:ApplicationName~ can be used for a
communication to a thread running on the same machine, and just \verb~ThreadName~
can be used to send to  another thread running within the same Qu-Prolog application. 
Thus, an \verb~ApplicationName~ is a thread name domain and a 
\verb~HostName~ is an application name domain. 

The special addresses \verb~self~ and \verb~creator~ respectively refer
to the thread itself, and the thread that forked it, providing there is such a thread. For a top level thread, creator denotes the thread. 

These addresses are used in this layer for sending messages and for
pattern matching against addresses in incoming messages.

The predicates
\begin{verbatim}
Message ->> Address
Message ->> Address reply_to ReplyAddress
\end{verbatim}
are used to {\em send} messages. So, for example,
\begin{verbatim}
connect ->> main_thread:server_process
\end{verbatim}
sends the \verb~connect~ symbol as a message to the thread with name \verb~main_thread~
in the Qu-Prolog application \verb~server_process~ on the local machine.  
\begin{verbatim}
connect ->> main_thread:server_process reply_to creator
\end{verbatim}
sends the same message but also sets its associated  reply-to address to the creator of the message sender.
Further examples of uses of communication using this layer are given later.

Each of the predicates
\begin{verbatim}
Message <<- Address
Message <<- Address reply_to ReplyAddress
\end{verbatim}
{\em reads the first message} from the message buffer and unifies it with
the supplied arguments. The call suspends if there
are no messages in the buffer. It fails if the first message does not unify with the supplied arguments.  

Each of the predicates
\begin{verbatim}
Message <<= Address
Message <<= Address reply_to ReplyAddress
\end{verbatim}
{\em searches} the message buffer looking for a message that unifies with the supplied
arguments. If one is found, that message is removed, otherwise the
call suspends until another message arrives. It continues, checking each newly 
 arrived message, until one does unify. 

The most powerful form of message receive is the {\tt message\_choice} call
which has
the form
\begin{verbatim}
message_choice (
    MsgGuard1 -> CallConj1
    ;
    MsgGuard2 -> CallConj2
    .
    .
    MsgGuardn -> CallConjn)
\end{verbatim}
where each {\tt MsgGuardi} has the form
\begin{verbatim}
    MsgPtn <<- S reply_to R :: Test
\end{verbatim}
in which the {\tt reply\_to R} and \verb~:: Test~ are optional.
This will scan the message buffer of the thread testing each message
against the sequence of alternative message guards in turn. When a
message is found which matches the {\tt MsgPtn <<- S reply\_to R}
of any one of the message guards, {\em and} the associated {\tt Test} call succeeds, the
message is removed from the buffer and the corresponding {\tt
CallConj} is executed. The {\tt Test} call can be an arbitrary Prolog
call that typically tests variable values generated by the
unification of the message with the message pattern 
\begin{verbatim}
MsgPtn << S reply_to R 
\end{verbatim}

As with the single message search operator {\tt <<=}, the {\tt message\_choice}
call will suspend if the end of the message buffer is reached and will be
automatically resumed 
when a new message is added.  However, in this case we can set a
limit on the time for which the {\tt message\_choice} call and the thread that executes it should
suspend. We do this by including
\begin{verbatim}
timeout(T) -> TimeOutCall
\end{verbatim}
as a  last alternative of the message choice call. This will limit to {\tt T} seconds the time that the call suspends, after the search for an acceptable message  has reached the
end of the buffer. When the time limit is reached the
{\tt message\_choice} call executes  {\tt TimeOutCall}. 

In the current implementation of the higher-level message communication operators, the communication options of the lower level communications predicates they invoke are set to do
encoding and to remember variable names. It is, however,
straightforward to modify
or extend the definitions if other behaviours are required. 
All the high level communications primitives are implemented as a library of 
Prolog programs that use the base level primitives.

\subsection{Local inter-thread communication using the dynamic database}
The only way that threads running in different Qu-Prolog applications can
communicate is via messages. 
However, Qu-Prolog threads running {\em within} a single Qu-Prolog application
can also communicate using the dynamic database.
When a thread
executes an {\tt assert} or {\tt record} this is added to the data
area accessible by all the local threads. An asserted clause can
later be accessed by another local thread using {\tt clause} or {\tt
retract}. By executing these calls as arguments to a special {\tt
thread\_wait} meta-call predicate, we can make the accessing call
suspend (rather than fail) until a matching clause is asserted by another local
thread. So threads within an application can synchronise either via
explicit message passing or by blocking accesses to the shared dynamic
memory. We shall use this second form of synchronisation for local threads 
in our Linda server example.

\section{Communications: Implementation}

Our original implementation of communications was based on ideas from April
and included a name server that kept track of symbolic names, and local caches
of mappings between symbolic address and low-level (TCP/IP) addresses.

At the same time Frank McCabe was developing the 
{\it InterAgent Communications Model} (ICM).
This model was developed from April and the newer versions of April use
this model for communications.

Recently, we replaced our communications support with the ICM API. 
We did this for a number of reasons. Firstly, it allowed us to concentrate more
on Qu-Prolog specific development rather than on communications development.
Secondly, the ICM is more robust than our original implementation with respect
to processes dying. Thirdly, the ICM has good support for such things as
`mobile computing' and lastly, it allows Qu-Prolog to communicate with other
applications, such as April applications, that use the ICM API.
We also found that it was very straightforward to add the ICM API to the Tk
interpreter, thereby making it easy to write GUI's that interact with
applications using message passing. It is then simple to produce a system that
has multiple Qu-Prolog applications interacting with multiple GUI's.

The ICM can be divided into two parts: the ICM communication servers that route messages 
from one process that uses  the ICM API to another such process; and the ICM API that provides
functions for connecting to and disconnecting from an ICM
communication  server and for
sending and receiving messages.

In order to use the ICM for communication, at least one ICM communication server needs to
be running on the network. Typically, for a wide area network there will be
one communication server per machine, but for a local area network there might be a single
ICM communication server running on a designated machine. However, for
simplicity of presentation, in  the rest of the paper we shall assume
that there is a communication server running on each machine on which
the Qu-Prolog application running. 

An application that wants to use the ICM registers its name with one of the 
ICM communication servers, typically the local communication server. The ICM
(if there is one) then takes responsibility for routing messages to and from this process.

ICM addresses are similar to the Qu-Prolog addresses described earlier. The
basic ICM address consists of a {\it home}, a {\it name} and a {\it target}.
The {\it home} is the name of a machine running an ICM communication server, the {\it name} is the name of a
process registered with this communication server, and the {\it target} is a field that is not used directly by the ICM, 
but is intended for use by the application. For Qu-Prolog the target is used as
the thread name (or ID).

When a named Qu-Prolog process is launched, the process registers its name
with the ICM communication server running on the same host (the
default). This opens two TCP connections, one for outgoing messages,
one for incoming messages. It then forks a POSIX thread for processing
the incoming
messages from the  communication server  and  begins execution of the initial Qu-Prolog thread.
The incoming message handling thread consists of a loop that waits for a message
from the communication server, decodes the message, and adds the message to the 
message buffer of the local thread identified  by the target field of the address.

The way outgoing message are handled depends on the recipient's address.
If the recipient is a thread with the same Qu-Prolog process then the message is 
simply copied to the recipient's message buffer.
Otherwise, the message is dispatched using functions from the ICM API
to the local communication server for routing to the target thread of
some other Qu-Prolog process. 
If the other Qu-Prolog process is
running on the same host, only the local ICM communication server is
involved. It looks up the name in its list of registered processes,
and sends the message via the TCP connection that was opened when the
process registered. If the Qu-Prolog process is on another host, the
local communication server dispatches the message to the communication
server running on that host, which is identified in the full thread address. To
do this, it may open a temporary TCP connection. The target communication server
then forwards it to the identified Qu-Prolog process, which, in turn,
puts it into the thread's message buffer. All this  middleware is invisible to the
Qu-Prolog application programmer. 

The ICM system has considerable
functionality for robust inter-host communication. For example, if the
target  Qu-Prolog process is
a registered process but is
temporarily down, its communication server will hold any
messages for any of its threads until the process resumes,
and re-connects with the communication server. It addition, for hosts
that may be  temporarily disconnected from the network, such as a
laptop computer, we can designate a proxy communication server that is
on a host permanently on the network. All messages for the communication
server on the laptop will then be automatically re-routed to the proxy
server when the laptop is disconnected. On re-connection, they will be
automatically downloaded to the laptop communication
server, for forwarding to its local processes and their
threads. Again, this is invisible to the Qu-Prolog application
programmer. 

\section{The Linda Model}

In this section we illustrate some of the multi-threading and high-level
communication features of Qu-Prolog by presenting an 
implementation of the Linda model
for inter-process communication \cite{LINDA}. 
Note however, that Qu-Prolog's primary form of communication is via message
passing,
not through the use of the Linda model or other forms of communication using
blackboards.

In the Linda model processes communicate by 
adding and removing data tuples to and from a 
shared tuple data space. 
They can suspend, waiting for a tuple that matches a certain pattern.

Each communication process can execute the  following operations
on the tuple space.

\begin{itemize}
\item {\tt out({\it Tuple})} -- Add {\it Tuple} to the tuple space.

\item {\tt in({\it Tuple})} -- Remove {\it Tuple} from the tuple space 
(block until match found).

\item {\tt rd({\it Tuple})} -- Lookup {\it Tuple} in the 
tuple space (block until match found).

\item {\tt inp({\it Tuple})} -- Remove {\it Tuple} from the tuple space 
(fail if match not found).

\item {\tt rdp({\it Tuple})} -- Lookup {\it Tuple} in the tuple space 
(fail if match not found).
\end{itemize}

For simplicity, we have chosen
not to deal with the Linda {\tt eval} operation. This could be implemented
using, for example, a variant of distributed querying given later.

Figure \ref{LS} presents an implementation of a Linda tuple 
server that uses the 
dynamic database
of a Qu-Prolog process, called {\tt linda\_server} when launched, to store the
tuple space. 
Each tuple of the tuple space becomes a  fact in the
dynamic database which is shared across all threads within
the {\tt linda\_server} process. Each thread can therefore access, assert
and retract any of these facts. 
The initial thread of this application, the one
started when the application is launched, is called {\tt
main\_linda\_thread}. Each process that wants to access the tuple
space must first register with the {\tt linda\_server} by sending a
{\tt connect} message to\\
{\tt main\_linda\_thread:linda\_server@hostname}. 

The server process is launched using the \verb~-A linda_server~ switch
which names the process. 
On startup, the initial server thread names itself 
and then enters a loop waiting for \verb~connect~ messages from a
client process. It ignores all other messages. 

\begin{figure}
\begin{verbatim}
main(_) :-
    thread_set_symbol(main_linda_thread), 
    linda_loop.

linda_loop :-
    repeat,
    connect <<= FromAddr, 
    thread_fork_anonymous(_, linda_thread(FromAddr)), 
    fail.    

linda_thread(A) :-
    connected ->> A, thread_loop(A). 

thread_loop(A) :-
    repeat, 
    message_choice (           % only from its client, A.
      out(T) <<- A -> assert(T), inserted ->> A
      ;
      in(T) <<- A -> thread_wait(retract(T)), ok(T) ->> A
      ;
      rd(T) <<- A -> thread_wait(clause(T, _)), ok(T) ->> A
      ;
      inp(T) <<- A -> (retract(T) -> ok(T) ->> A ; fail ->> A)
      ;
      rdp(T) <<- A -> (clause(T, _) -> ok(T) ->> A ; fail ->> A)
    ),
    fail.
\end{verbatim}
\caption{The Linda Server}
\label{LS}
\end{figure}
On receipt of a \verb~connect~ message 
the process forks a thread to deal with the client. The created
thread sends a \verb~connected ~acknowledgement to the client, which
also serves to identify the thread to the client (as the sender of the
acknowledgement), and then enters a
loop to process client requests.
The thread will suspend waiting for
the client's requests. It processes them by appropriate operations on the
dynamic database. The
multi-threading of Qu-Prolog allows a very simple and elegant
implementation of the Linda model. The client processes of the tuple
space manager can be
distributed over the Internet. 

The predicate \verb~thread_wait~/1 causes the thread to block until
the supplied tuple term {\tt T} gets asserted.
Note that by using one thread per client, only those threads that should
block do so and the server can continue to process commands from
(non-blocked) clients.

The code in Figure \ref{LCS} provides a library of predicates for use by a 
Linda
client implemented in Qu-Prolog.
For this example we assume \verb~linda_machine~ is the machine on which
a Linda server is running.

\begin{figure}
\begin{verbatim}
remember_linda_thread_address(A) :-
    my_id(ID), assert(linda_th_addr(ID,A)).
get_linda_thread_address(A) :-
    my_id(ID), linda_th_addr(ID,A).

linda_connect :-
    connect >> main_linda_thread : linda_server @ linda_machine,
    connected <<= A,
    remember_linda_thread_address(A).
linda_disconnect :-
    thread_tid(TID),
    retract(linda_th_addr(TID,A)),
    disconnect ->> A.

linda_out(T) :-
    get_linda_address(A), out(T) ->> A, inserted <<= A.
linda_in(T) :-
    get_linda_address(A), in(T) ->> A, ok(T) <<= A.
linda_rd(T) :-
    get_linda_address(A), rd(T) ->> A, ok(T) <<= A.
linda_inp(T) :-
    get_linda_address(A), inp(T) ->> A, M <<= A, M = ok(T).
linda_rdp(T) :-
    get_linda_address(A), rdp(T) ->> A, M <<= A, M = ok(T).
\end{verbatim}
\caption{Linda Client Support}
\label{LCS}
\end{figure}

Each Linda client communicates with a Linda thread created specifically to 
handle this client. Each client therefore needs to keep track of the
address of its Linda thread. We have chosen to do this by asserting
the address together with the client thread identifier. 
The client thread identifier is included to avoid confusion when
several Linda clients are running in one Qu-Prolog process and therefore sharing the
dynamic database.

Note in particular the use of symbolic names to identify the
destination in the \verb~connect~ message send,
and the  use by the client of the
the identity of the sender of the {\tt connected} reply it receives as
the identity of its server thread.

The blocking behaviour of the operations is achieved on the client's side
by message blocking and on the server's side by the use of 
\verb~thread_wait~/1.  The clients need not be Qu-Prolog
threads. Using the ICM API they could, for example,  be C or April
processes using Qu-Prolog as the Linda tuple space handler.

\section{Distributed querying}
We have seen that it is straightforward to have messages sent between
Qu-Prolog threads running in Qu-Prolog applications anywhere on the
internet. Let us consider an application in which we have several
Qu-Prologs running, one per host, each of which has its own 'deductive database' of
Prolog rules. Imagine a query interface that allows a user to enter
a query on any host to any one of these query servers, which may be on
another host.  Imagine the interface allows the querier to request that either
{\em all} the answers be returned, in one reply, or that the answers
should be returned 
{\em one at
a time}, on demand. 

To handle the query requests, each remote query can have a main thread  that accepts either an {\tt all\_of(C)}
message -  a request to the server to produce all solutions of the
query {\tt C} - or a {\tt stream\_of(C)} message -
a request to produce the  answers  one at a time.  It handles the
former by using findall to construct the reply.
It  handles the latter by forking a temporary thread to 
 interact with the user, who can request the answers one at
a time, or terminate the query thread at any stage. Forking a
temporary thread  allows the main query
thread  to deal  with other queries, from the same or other users of
the distributed information system.

The top level of the query server is similar to the Linda server
and is presented in Figure \ref{QS}. When
launched with a \verb~-A query_server~ switch, 
the query server's main thread starts executing and names itself 
\verb~query_thread~ .   
A client can then interact with the server by sending a query to
\verb~query_thread:query_server@machine~. 

\begin{figure}
\begin{verbatim} 
main(_) :-
    thread_set_symbol(query_thread), 
    query_loop.

query_loop :-
    repeat, 
    message_choice  (
      all_of(Call) <<- _ reply_to R -> findall(Call,Call,L),
                                   answer_list(L) ->> R
      ;
      stream_of(Call) <<- _ reply_to R -> 
                             thread_fork_anonymous(I,ans_gen(Call,R)),
                             query_thread_is(I) ->> R
    ),
    fail.

ans_gen(Call,R):-
   call(Call),
   answer_instance(Call) ->> R, % send answer to client R
   message_choice (
      next <<- R -> fail   % fail back to get next answer
      ;
      finish <<- R -> thread_exit). % terminate on finish
ans_gen(Call,R) :-
   fail ->> R,    % when no more answers send fail to client 
   thread_exit.  % and terminate
\end{verbatim}
\caption{A Query Server}
\label{QS}
\end{figure}

Note that an {\tt all\_of} message is handled by executing a {\tt findall}
within
the main query thread. However, the {\tt stream\_of} message causes a new
thread to be started, executing the call {\tt ans\_gen(Call,R)}
where {\tt Call} is the
query and {\tt R} is the thread to which answers will be sent - the
client. Notice that this time the identity of the  forked query thread is sent to the client
in a {\tt query\_thread\_is} message from the main
thread. 
Alternatively, as in the Linda program,  we could have made
the temporary query thread identify itself to the client with an
initial message. The client will now interact with this
temporary thread. 
Notice that the server identifies the client as the
\verb~reply_to~ of the received query, rather than the \verb~sender~
of the query.  This allows queries to be forwarded to a query server
on behalf of another thread.  Such forwarding might be used by broker
Qu-Prolog applications 
acting as intermediaries in the distributed information system.

The {\tt ans\_gen}  program
finds the first  solution instance of  {\tt Call}, sends it to the
client {\tt R}, and then
waits for a {\tt next} or {\tt finish} message to arrive from {\tt R} to
see if it should find another solution or not. If, when it has
received a {\tt next} message, there are no more
solutions, it signals this by replying with the message {\tt fail}. 
A client can also send a {\tt finish} message, to prematurely terminate the search
for solutions. 

Figure \ref{QSCS} presents an interface to a client program interface
to any of the query servers via
two meta-call predicates {\tt ?} and {\tt ??}.
A call of the form \verb~C?QS~  sends an {\tt all\_of(C)}
message to the query server {\tt QS}, waits for the list of solutions
from the server, and then uses {\tt member} to locally backtrack over
the solutions as required.

A call  of the form \verb~C??QS~ sends a {\tt stream\_of(C)}
message to {\tt QS}, waits for the thread ID and the first answer, and then uses
{\tt deal\_with\_ans} to process the reply and to manage any
subsequent backtracking.

\begin{figure}
\begin{verbatim}
Call?Q_S:-
  all_of(Call) ->> Q_S,
  answer_list(L) <<= Q_S,
  member(Call,L).

Call??Q_S :-
  stream_of(Call) ->> Q_S,  % send stream_of query to Q_S
  query_thread_is(QTh) <<= Q_S, % wait for id of new thread
  Ans <<= QTh,  % wait for first answer from this thread
  deal_with_ans(Ans,Call,QTh).

deal_with_ans(fail,_,QTh) :- % answer is fail message
   !, fail.                  % no (more) answers
deal_with_ans(answer_instance(C),C,_).
deal_with_ans(answer_instance(_),C,QTh):- 
  next ->> QTh, % request next ans for remote C call
  Ans <<= QTh,
  deal_with_ans(Ans,C,QTh).
\end{verbatim}
\caption{Query Server Client Support}
\label{QSCS}
\end{figure}
The above meta calls can be used in user queries and in clauses in the
databases of each query server. So a user query to one server can result
in a chain of remote queries being sent over the network of query
servers. The initial query server thus serves as an interface to the
entire network.

There is a slight problem with the above implementation of remote querying. 
If a query or clause  executes a cut ({\tt !})
after a {\tt C??QS} call, but before all the solutions have been requested 
and
returned, the temporary thread created by {\tt QS} will not be exited, and
will be left as an orphan. A
slight elaboration of the query server program and the {\tt ??} program will ensure
that such orphan threads are all sent a \verb~finish~ message, providing the distributed query evaluation is started with a top level user query to one of the servers.
We will not give the code, but we will
explain the idea.

The user interface program, and each query server thread  (even a temporary thread), remembers
the identities of all the remote query threads  started as a result of
a {\tt ??} call it executes. A thread {\tt T} forgets the identity of one of these remote query threads 
{\tt QTh} if {\tt QTh} indicates its termination by sending 
{\tt T} a {\tt fail} message. The remembering and forgetting  can be done  by having the {\tt ??}
program assert a {\tt remote\_thread(T,QTh)}  fact when it gets the 
{\tt query\_thread\_is(QTh)}
message, and having {\tt deal\_with\_ans} retract the fact on  receipt of the
{\tt fail} message from {\tt QTh}, indicating its normal termination.
  
We now modify the \verb~query_loop~ program so that it executes
a call to:
\begin{verbatim}
kill_orphans :-
   my_id(ID), 
   forall(retract(remote_thread(ID, QTh)),finish >> QTh).
\end{verbatim}
after it has found and returned all the solutions to an {\tt all\_of} query  request.  This will cause all remote query threads started during its execution to be terminated. In addition, we modify the
  program for {\tt ans\_gen} so that it calls {\tt kill\_orphans} just
  before it executes a {\tt thread\_exit}, both on normal termination
  (after it has returned all its answers) and
  on receipt of a {\tt finish} message (premature termination). The
  latter, which is a propagation of {\tt finish} messages, implements
  distributed garbage collection of query threads started when a {\tt finish} is sent  either by the user interface program, or when any query thread has found all the solutions to its query.

A {\tt finish} will be sent by the user interface if the user indicated one at a time answers causing the query to be dispatched as a {\tt stream\_of} request, 
 {\em and} the user indicates they want no more answers before all the answers
 have been returned (equivalent to a top level !).  The user interface  initiates
  distributed garbage  collection of what would be orphan threads by sending a \verb~finish~
message to the temporary query thread handling its {\tt stream\_of} request.
On receipt of the {\tt finish}, this thread
will, in turn,  execute  {\tt kill\_orphans} and so send \verb~finish~
messages to any remote threads it
started, that have not yet terminated.  They, in turn, will execute
 {\tt kill\_orphans}, effectively forwarding the \verb~finish~
 message to their orphan threads. Eventually, all orphaned query
 threads started by the user query, directly or indirectly,   will be terminated. 

The other case to consider is when all the answers to the user query have been  returned to the user
 interface program.  Whether or not the user query was dispatched as an 
{\tt all\_of} or
a {\tt stream\_of} remote query, garbage collection of the orphan threads will be
  started by the query thread that handled the query  by it calling
 {\tt kill\_orphans} when all the answers have been returned to the user interface. 

\section{Comparisons}

In this section we briefly compare communication in 
Qu-Prolog with that of SICStus-MT, BinProlog, CIAO, Erlang, Mozart-Oz and April.

For messages between threads within the same Prolog process, SICStus-MT
uses
much the same approach as Qu-Prolog - both have a single message buffer,
which they call a port, 
both are able to scan the buffer looking for message patterns, and both
suspend if no (matching) messages are found.
The main differences are that SICStus-MT  does not
use message buffers for 
communication between threads in {\em different} SICStus-MT processes,
and symbolic names are not used for threads. To communicate between
different Prolog processes TCP communication primitives must be used.

The main method of high-level communication used by BinProlog is 
through the
use of Linda tuple spaces.
Our implementation of the Linda model demonstrates that it is easy to
emulate the BinProlog style in Qu-Prolog. On the other hand, 
it would also be easy to emulate the Qu-Prolog style of communication
using BinProlog's tuple space. The symbolic addresses could be included
as extra arguments to tuples stored in the tuple space and this 
information could be used by threads looking for messages meant for them.

If efficiency of communication is measured by the number of communications
needed to send a message from source to destination then a comparison
can be made between the two systems. Assuming, in the BinProlog system,
one tuple space is used then three communications are required: one to
put the message in the tuple space; one to ask the tuple space for
a message; and one for the tuple space to send the message.
In Qu-Prolog the number required depends on the number of ICM 
communication servers `between' the sender and receiver. 
If the sender, receiver  are in Qu-Prolog process registed with the
same ICM communciation server 
then two communications are required, if they are registered with
different commuication servers then three communications are required. 
Note that when messages are sent between threads in the
same Qu-Prolog process then no communications are required - term copying from the 
heap of the sender to the buffer 
of  the receiver is
used.

In the case of Qu-Prolog the other overheads of communication 
relate to the message handling thread - ICM message decoding and copying to 
message buffers.
In BinProlog the main extra overheads seem to be related to the management
of the tuple space.

The CIAO system \cite{CIAODB}  uses of the dynamic Prolog
database for  
communicating between threads in the same process.   Whereas Qu-Prolog uses
{\tt assert} and {\tt retract} to update the database and \verb~thread_wait~
to suspend calls to the database, the CIAO system uses extensions of
the normal dynamic  
database access/update predicates that automatically suspend if a
thread tries to access a clause for a dynamic predicate declared as
\verb~concurrent~. The concurrent predicates are the ones that are
used for inter-thread communication.  This automatic suspension also
applies to normal calls to a concurrent predicate, even on
backtracking. Thus, a thread will suspend when a call to a concurrent
predicate has `seen' all the clauses for the
predicate that have so far been asserted by the other threads. This
allows the dynamic database to be used to communicate a stream of
data between threads, as an incrementally asserted set of facts, with
automatic suspension of  consuming threads that run ahead of the
producers. In Qu-Prolog we would normally achieve this by using
message communication between the consumers and a single  intermediary thread
that multicasts each data item to all the consumers.  

The CIAO inter-thread  communication  can be enhanced by the use of 
attributed variables \cite{Hm} to allow communication  of variable
bindings between threads via the dynamic database.  Attributes are
terms that can be associated with unbound variables.  Whenever a variable with
an attached attribute is bound, a user defined program is
automatically invoked that
is passed the variables current attribute values  and the value to which it is bound. The
attribute value(s)  can be used to uniquely identify the
variable. The invoked user defined program can
then assert the attribute/value pair in the dynamic database, thereby
making the binding available to other threads. \cite{Hm} shows how this
mechanism can be used to implement concurrent processes apparently
communicating  via
incrementally generated bindings for shared variables, when the only
communication is via the dynamic database.   

We have not previously mentioned this, but variables in Qu-Prolog can have delayed goals associated
with them that are woken when the variable becomes instantiated (even to 
another variable).  
Following \cite{Hm}, we believe we might be able to use this mechanism
and the \verb~remember_names~ feature of Qu-Prolog to also implement `shared
variable' communication between Qu-Prolog local threads.

Erlang is essentially a concurrent committed choice
logic programming language with a functional  syntax. However, instead
of communicating between the different processes implicitly, via
incrementally generated bindings of shared variables, explicit
communication via message send and receive operations are used. As with
Qu-Prolog, each Erlang process has a single message buffer, and
messages are read from  the buffer using a disjunction of guarded
commands,  very like the Qu-Prolog \verb~message_choice~ operator. In
fact, 
Qu-Prolog's 
\verb~message_choice~ operator  is modelled on the disjunctive message receive of
Erlang. Erlang processes can only communicate via messages, there is
no shared database. In addition the language has only  pattern
matching, not unification.

Mozart-Oz \cite{Har} is essentially a concurrent constraint programming
language extended to support Prolog style query evaluation, objects, 
and  communication over networks.  The primary form of communication
between Mozart processes, which must be explicitly launched, is
via shared data stores that can hold any Mozart value, including unbound
variables. When a value {\tt V} is posted to such a store, which a process must
do explicitly by calling a special system method with {\tt V} as
argument, the store returns a ticket {\tt T} uniquely identifying {\tt V}
in the local store. This is
an ASCII string that is an internet wide unique identity for {\tt V}.  (It includes the identity of the host which holds the store as
well as a store unique identity for {\tt V}.)   Any other Mozart process, whether local or remote, can retrieve the
value by calling another system method with the ticket {\tt T} as argument.
In addition, if the posted value is an unbound variable, any number of
other  processes
can invoke another system method to set a watch on the variable,
again identifying the variable by its
ticket.  Then, when the variable is bound, by the process that posted
it to the store, or any other process that has gained normal access to it using its ticket, all the watch
processes will be automatically
sent its value. This gives a multi-cast mechanism via shared variables
placed in stores.  
 
April is not a logic programming language, it is a
higher-order (in the functional programming sense) 
distributed symbolic language. However, it is 
similar to both Erlang and  Qu-Prolog in that each April process has a
single message buffer from which messages can be extracted using a
disjunction of guarded message receives containing message patterns.  April and Qu-Prolog
both use the ICM message transport system. April messages must be
completed ground terms but higher order values, such as
function and procedure closures, as well as objects (similar to Mozart objects), can be sent in
messages.  Both Erlang and April influenced the design of the
inter-thread communication of Qu-Prolog.   

\section{Conclusion}

In this paper we presented the multi-threading and message communication
capabilities of Qu-Prolog and outlined the implementation
of message processing.
The high-level methods for sending and receiving messages  were
discussed and examples of the implementations of the Linda model
and a distributed query server system were presented. 

We have also  implemented a
concurrent OO extension of Qu-Prolog in which objects are active and
are each executing  as separate
threads.   Each active object acts rather like a query server,
responding to calls on its clauses which are sent as messages from
other active objects. These clauses are the
methods  of the object.  They are given in class definitions that can use
multiple inheritance to define the method clauses of a given
class. Predicate definitions within each class hierarchy are disjoint,
even if they use overlapping predicate names. This is implemented
using predicate renaming, invisible to the programmer.  State for each active
object can be represented as property values stored in the record data
base, or as clauses for special dynamic predicates, declared as state predicates
of the object's class. Asserted clauses for such dynamic predicates always
include the identity of the object (thread) that asserted them, so
state clauses for the different objects of the same class, even when running in
the same Qu-Prolog,  are distinguishable.
This implicit indexing of the dynamic clause of an object by its identity is invisible to the
programmer.  Another key feature is a special system predicate, \verb~my_state/1~, that allows all the dynamic clauses
and property values of an executing  object \verb~O~, capturing its
current state, to be reified as a
list. This can then be sent in a message to an object server, that can
use it to
launch an object with the same state on another machine. This can be
an object of the same class as \verb~O~ providing
that the new machine has access to the class definition for
\verb~O~, which could even be fetched from a code server. This gives us object cloning, and allows us to program
applications with
mobile agents implemented as active deductive objects that
transport themselves in this way.

Our thesis is that the combination of multiple threads and high-level
communication using symbolic addresses supported by Qu-Prolog
provides application writers with simple and powerful techniques
for implementing a wide range of intelligent distributed
systems, possibly opening up new application areas for logic
programming.


\begin{thebibliography}{}


\bibitem[\protect\citename{Armstrong {\em et~al.}\relax, }1993]{Erlang}
J. Armstrong and R. Virding and M. Williams.
\newblock {\em Concurrent Programming in Erlang}.
Prentice-Hall, 1993.

\bibitem[\protect\citename{Becht {\em et~al.}\relax, }1996]{ERGO} 
Holger Becht, Anthony Bloesch, Ray Nickson and Mark Utting. 
\newblock Ergo 4.1 Reference Manual,
\newblock Technical Report No. 96-31,
Software Verification Research Centre,
University of Queensland.

\bibitem[\protect\citename{Carriero {\em et~al.}\relax, }1989]{LINDA}
N. Carriero and D. Gelernter
\newblock Linda in context.
\newblock {\it CACM}, 32(4),1989, pp. 444--458.


\bibitem[\protect\citename{Carrington {\em et~al.}\relax, }1996]{PREF} 
David Carrington, Ian Hayes, Ray Nickson, Geoffrey Watson and Jim Welsh,
\newblock A Tool for Developing Correct Programs by Refinement.
\newblock {\it Proc.\ BCS 7th Refinement Workshop, Bath, UK},
(ed. He Jifeng),
Electronic Workshops in Computing,
Springer, 1996, pp. 1--17.

\bibitem[\protect\citename{Carro {\em et~al.}\relax, }1999]{CIAODB}
Manual Carro and Manuel Hermenegildo,
\newblock Concurrency in Prolog Using Threads and a Shared Database.
\newblock {\it Proceedings of ICLP99}, (ed. D. De Schreye),
\newblock MIT Press, 1999, pp. 320--334.

\bibitem[\protect\citename{Clark {\em et~al.}\relax, }1998]{DAI}
Keith Clark, Peter J. Robinson and Richard Hagen,
\newblock Programming Internet Based DAI Applications in Qu-Prolog,
\newblock {\em Multi-agent systems}, (eds C. Zhang, D. Lukose),
\newblock Springer-Verlag LNAI 1544, 1998, pp. 137--151.

\bibitem[\protect\citename{Cook {\em et~al.}\relax, }1999]{MTTP}
Phil Cook and Peter~J. Robinson.
\newblock Multi-threading in an interactive theorem prover.
\newblock Technical Report No. 99-01,
Software Verification Research Centre,
University of Queensland.

\bibitem[\protect\citename{Eskilson {\em et~al.}\relax, }1998]{SICSMT}
Jesper Eskilson and Mats Carlsson.
\newblock SICStus MT - A Multithreaded Execution Environment for SICStus
Prolog.
\newblock {\em Principles of Declarative Programming},
(eds. Catuscia Palamidessi, Hugh Glaser, Karl Meinke),
Springer-Verlag LNCS 1490, 1998, pp. 36--53.

\bibitem[\protect\citename{Fidge {\em et~al.}\relax, }1995]{VSOFT}
C. Fidge, P. Kearney and M. Utting.
\newblock Interactively Verifying a Simple Real-time Scheduler.
\newblock In P. Wolper, editor, {\it
Computer Aided Verification},
Springer-Verlag LNCS 939, pp. 395--408, 1995.

\bibitem[\protect\citename{Hagen {\em et~al.}\relax, }1999]{QPRM}
Richard~A. Hagen and Peter~J. Robinson.
\newblock Qu-Prolog 4.3 Reference Manual.
\newblock Technical Report No. 99-03,
Software Verification Research Centre,
University of Queensland.

\bibitem[\protect\citename{Haridi {\em et~al.}\relax, }1998]{Har}
S. Haridi, P. von Roy, P. Brand and C. Schulte,
\newblock Programming languages for distributed applications.
\newblock {\em New Generation Computing},
\newblock 16(3), 1998, pp. 223--261.

\bibitem[\protect\citename{Hemenegildo {\em et~al.}\relax, }1995]{Hm}
M. Hemenegildo, D. Cabenza and M. Carro.
\newblock On the uses of Attributed Variables in Parallel and
Concurrent Logic Programming Systems
\newblock In L Sterling, editor, {\it Proceedings of ICLP95}, MIT
Press, 1995, pp. 631--645.

\bibitem[\protect\citename{McCabe {\em et~al.}\relax, }1995]{ClMc}
F. G. McCabe and K. L. Clark.
\newblock April:Agent Process Interaction Language.
\newblock {\em Intelligent Agents},
(ed. N. Jennings, M. Wooldridge), Springer-Verlag LNCS 890, 1995, pp. 324--340.

\bibitem[\protect\citename{McCabe\relax, }1999]{Mc}
F. G. McCabe
\newblock ICM Reference Manual.
\newblock Fujitsu Labs of America,
\newblock http://www.nar.fla.com/icm/manual.html, 1999.

\bibitem[\protect\citename{Tarau\relax, }1997]{BP}
Paul Tarau.
\newblock BinProlog 5.75 User Guide.
\newblock Technical Report 97-1, D\'epartment d'Informatique,
Universit\'e de Moncton, April 1997.

\bibitem[\protect\citename{Robinson\relax, }1997]{QPUM}
Peter~J. Robinson.
\newblock Qu-Prolog 4.2 User Guide.
\newblock Technical Report No. 97-12,
Software Verification Research Centre,
University of Queensland.

\end{thebibliography}
\end{document}